\documentclass[twocolumn]{aastex62}
\usepackage{chngcntr}
\usepackage{booktabs}
\usepackage{xspace}
\usepackage{gensymb}

\newcommand{\msun}{$\mathrm{M_{\odot}}$\xspace}

\counterwithout{figure}{section}

\graphicspath{ {./figures/} }
\received{7 July 2020}
\revised{23 July 2020}
\accepted{30 July 2020}
\submitjournal{ApJ}

\shorttitle{Kinematic Evidence of GI}
\shortauthors{Hall et al.}

\begin{document}

\title{Predicting the kinematic evidence of gravitational instability}

\correspondingauthor{C. Hall}
\email{cassandra.hall@uga.edu}

\author[0000-0002-8138-0425]{C. Hall}
\affil{School of Physics \& Astronomy, University of Leicester, University Road, Leicester, LE1 7RH, U.K.}
\affil{Department of Physics and Astronomy, The University of Georgia, Athens, GA 30602, USA.} 
\affil{Center for Simulational Physics, The University of Georgia, Athens, GA 30602, USA.}
%\affiliation{Winton Fellow}

\author[0000-0001-9290-7846]{R. Dong}
\affiliation{Department of Physics \& Astronomy, University of Victoria, Victoria BC V8P 1A1, Canada}

\author[0000-0003-1534-5186]{R. Teague}
\affil{Center for Astrophysics | Harvard \& Smithsonian, 60 Garden Street, Cambridge, MA 02138, USA}

\author[0000-0002-8590-7271]{J. Terry}
\affil{Department of Physics and Astronomy, The University of Georgia, Athens, GA 30602, USA.}

\author[0000-0001-5907-5179]{C. Pinte}
 \affil{School of Physics and Astronomy, Monash University, Clayton Vic 3800, Australia}
 \affil{Univ. Grenoble Alpes, CNRS, IPAG, F-38000 Grenoble, France}

\author{T. Paneque-Carre\~no}
\affil{Departamento de Astronomica, Universidad de Chile, Camino El Observatorio 1515, Las Condes, Santiago, Chile}

\author{B. Veronesi}
\affil{Dipartimento di Fisica, Universita degli Studi di Milano, Via Celoria, 16, Milano, I-20133, Italy}

\author[0000-0001-6410-2899]{R. D. Alexander}
\affil{School of Physics \& Astronomy, University of Leicester, University Road, Leicester, LE1 7RH, U.K.}

\author[0000-0002-2357-7692]{G. Lodato}
\affil{Dipartimento di Fisica, Universita degli Studi di Milano, Via Celoria, 16, Milano, I-20133, Italy}

%\author{Ken Rice}
%\affiliation{IfA, Edinburgh}

\begin{abstract}

Observations with the Atacama Large Millimeter/Submillimeter array (ALMA) have dramatically improved our understanding of the site of exoplanet formation: protoplanetary discs. However, many basic properties of these discs are not well-understood. The most fundamental of these is the total disc mass, which sets the mass budget for planet formation. Discs with sufficiently high masses can excite gravitational instability and drive spiral arms that are detectable with ALMA %  \citep{dipierroetal2014,dipierroetal2015,halletal2016, halletal2019}
. Although spirals have been detected in ALMA observations of the dust %  \citep{perezetal2016,dsharp,dsharpspirals}
, their association with gravitational instability, and high disc masses, is far from clear%  \citep{meruetal2017,tomidaetal2017,halletal2018,forganetal2018}
. Here we report a prediction for kinematic evidence of gravitational instability. Using hydrodynamics simulations coupled with radiative transfer calculations, we show that a disc undergoing such instability has clear kinematic signatures in molecular line observations across the entire disc azimuth and radius which are independent of viewing angle. If these signatures are detected, it will provide the clearest evidence for the occurrence of gravitational instability in planet-forming discs, and provide a crucial way to measure disc masses.\\ ~\\

\end{abstract}

\section{Introduction}
It has become clear that most, if not all, protoplanetary discs contain some degree of substructure. Specifically, spirals have been readily observed in discs in scattered light at micron wavelengths   \citep{benistyetal2015,stolkeretal2016} and in dust emission at mm wavelengths   \citep{perezetal2016,dongetal2018,dsharpspirals}. Unlike other structures, such as rings, which are readily explained by planets   \citep{dipierro2015,dipierroetal2018} thanks to kinematic detections \citep{teagueetal2018,pinteetal2018,teagueetal2019a,pintenature,pinteetal2020}, thermal detections \citep{keppleretal2018} and accretion confirmation \citep{haffertetal2019}, the origin of spiral morphology remains ambiguous.

%Two main hypotheses have been proposed to explain spirals in general, in addition to alternatives such as binary and flybys (Price et al. 2018; Cuello et al. 2019) that may function in special cases. 
Density waves excited by $\gtrsim$ Jovian mass planets can quantitatively match the observed spirals in scattered light in both contrast and morphology  \citep{dongetal2015,fungdong2015,dongfung2017}. Some spirals may be due to binary companions, either internal to the disc  \citep{priceetal2018} or external to it   \citep{forganetal2018}. While possible, it is unlikely that most spirals are caused by stellar flybys since close encounters between stars are statistically much rarer in the majority of star formation regions compared with the observed occurrence rate of spirals   \citep{winteretal2018}.

Meanwhile, gravitational instability (GI) can also produce spirals. 
%GI occurs in protoplanetary discs when the mass of the disc is comparable to the mass of the central star. 
As a rule of thumb, a disc-to-star mass ratio $\gtrsim10\%$ is needed to trigger GI and produce detectable spirals  \citep{donghallrice2015,halletal2016,halletal2019}. 

However, directly measuring disc mass is almost impossible. The main constituent, molecular hydrogen (H$_2$), lacks a dipole moment, and at the low temperatures found in the bulk of protoplanetary discs emits only through faint quadrupole transitions. Disc masses can, however, be estimated by converting continuum flux density at millimeter/sub-millimeter wavelengths to a total dust mass, then scaling by a constant ratio to obtain a total gas mass   \citep{beckwithetal1990}. This method is plagued by uncertainties in basic quantities such as the dust opacity and the dust to gas mass ratio   \citep{andrews2020}. A method thought to be more accurate is measurement of line emission from other molecules thought to trace H$_2$, including HD   \citep{berginetal2013,mcclureetal2016}, CO, and its less abundant isotopologues   \citep{williamsbest2014}, and converted to H$_2$ mass through assumed abundance ratios. However, molecular abundances are believed to vary both spatially and temporally within a disc  \citep{ileeetal2017,quenardetal2018,zhangetal2019}, rendering the conversion from measured line flux density to a total gas mass highly model dependent  \citep{trapmanetal2017}.

Differentiating between these hypotheses is of crucial importance in planet formation. If spirals are predominately produced by giant planets, then the detection rate of such spirals can directly inform us about the occurrence rate and properties of giant planets  \citep{halletal2017,forganhallmerurice2018,dongnajitabrittain2018}. If, on the other hand, most spirals are caused by GI, their existence and morphology can be used to infer fundamental disc properties, such as disc mass   \citep{donghallrice2015,haworthetal2020,cadmanetal2020}, and therefore constrain planetary mass budgets and formation timescales \citep{nayakshinetal2020} on a comprehensive scale.

However, identifying the true origin of spirals in discs is difficult. The best way to confirm the planetary origin is to directly detect the putative spiral-causing planets. Except in rare cases   \citep{wagneretal2019}, planets associated with spiral structures have largely evaded detection in direct imaging searches to date, possibly because they are faint   \citep{brittainetal2020,humphriesetal2020}. 

%CH: removed this paragraph for brevity 
%The two hypotheses also make distinct predictions on the pattern speeds of spirals, with the planet-induced ones corotate with the driver and the GI-induced ones moving at local Keplerian speed. However, spiral pattern speeds are generally low, making their measurements difficult (Ren et al. 2018).

Simulations have shown that it is theoretically possible to differentiate between GI and planets by measurement of spiral pitch angles. GI creates logarithmic, symmetric spiral arms   \citep{halletal2016,forganetal2018}, with the number of spiral arms, $m$, determined by the disc-to-star mass ratio such that $m \sim 1/q$   \citep{donghallrice2015}. 
%RD: I wonder whether this could be moved forward to provide an example to the statement "(spiral) morphology can be used to infer fundamental disc properties, such as disc mass". 
Planets and external binary companions, instead induce spirals with variable pitch angles   \citep{dongetal2015,forganetal2018}. However, in practice, synthesised observations of spirals formed through these two mechanisms, at both mm and micron wavelengths, appear very similar with current instrumentation   \citep{donghallrice2015,dongetal2015,meruetal2017}.

\begin{figure*}[t!]
\centering
  \includegraphics[width=0.85\textwidth]{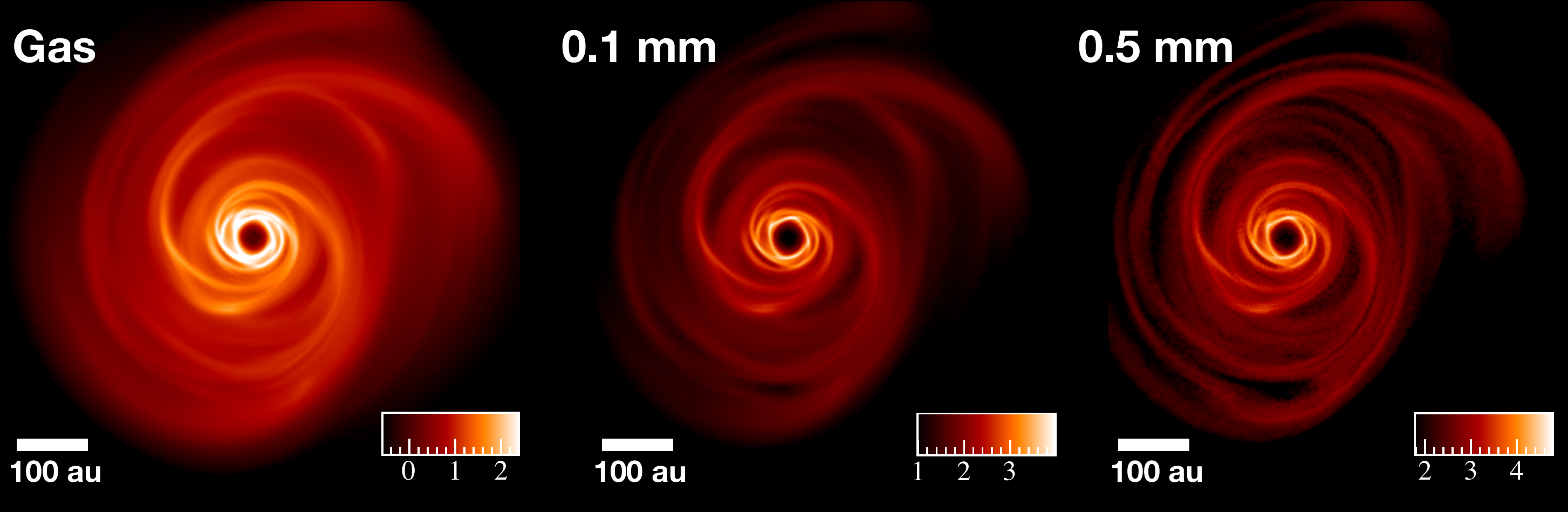}
\caption{\textbf{Hydrodynamical model of a self-gravitating disc.} Rendered surface density images of gas and dust grains of varying sizes, in units of log$_{10}$ g cm$^{-2}$. Small dust grains are well-mixed with the gas, while dust grains up to $\sim$5mm experience particle trapping. The maximal radial extent of each disc is $R\sim 300$ au. \label{fig:splash}}
%\end{figure*}
%\begin{figure*}
\vspace{0.25cm}
  \includegraphics[width=\textwidth]{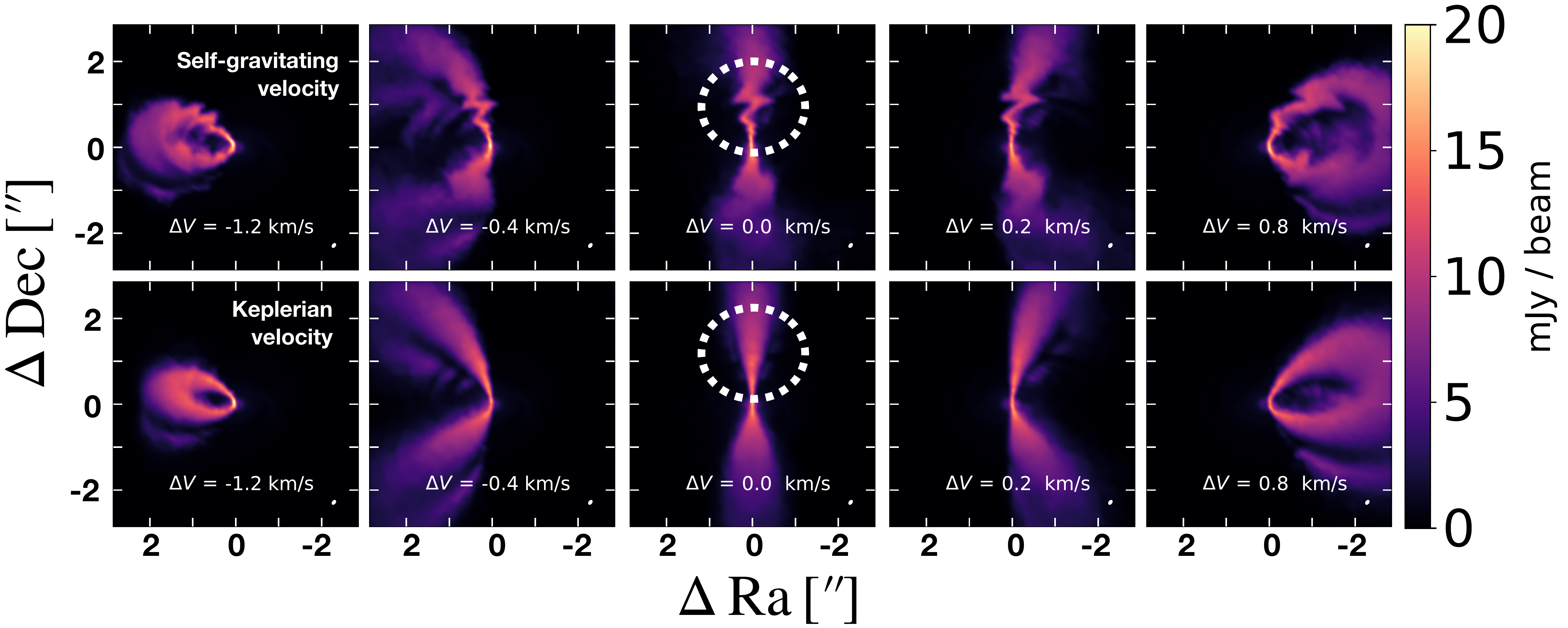}
\caption{\textbf{Predicted emission for gravitational instability.} Hydrodynamical model of a self-gravitating disc, post-processed with radiative transfer to produce emission maps of $^{13}$CO $J=3\rightarrow 2$ transition. Top row shows the unaltered, self-gravitating velocity structure, while the bottom row assumes a purely Keplerian velocity. The ``GI wiggle'' is circled in the top central channel, and is visible at all deviations from the systemic velocity. The bottom row shows this signature is not present if the velocities are not self-gravitating. \label{fig:channelmaps}}
\end{figure*}
%RD: one potential thing that might confuse readers who only scan figures is the white circle in the middle panel. One might somehow think only this feature is the target, and similar features in other channels are not. Technically the GI feature is present in all panels. Perhaps some clarificatin could be done to avoid confusion.

Our approach in this work is to present the dynamical effect of GI on the gas disc, traceable with molecular line emission, and introduce new diagnostic tools to identify GI unstable discs with gas observations. In the planet-in-disc case, the bulk kinematics are dominated by Keplerian rotation, modulated by a radial pressure gradient such that the gas orbits at slightly sub-Keplerian velocity. If a planet is present, then its wake will cause a deviation that is strongest closest to the planet, resulting in a localised ``kink'' near the planet when observed in molecular line emission \citep{perezetal2015,teagueetal2018,pinteetal2018,perezetal2018,teagueetal2019a,pintenature,pinteetal2020}.

\section{Methods}

\subsection{Hydrodynamical model.}
\label{subsec:hydro}
We performed a three-dimensional, dusty, gaseous global hydrodynamical simulation using the Phantom Smoothed Particle Hydrodynamics code  \citep{phantom}. Dust was modelled self-consistently with the gas using the ``one-fluid'' technique \citep{multigrain,ballabioetal2018} in the strongly-coupled regime. We used 1 million SPH particles, and followed the grain fraction of dust particles in sizes ranging from 1 micron to 4 mm in 5 size bins. We included dust since the temperature of dust sets the thermal structure for the surrounding gas, and we also include the force exerted on the gas by the dust, since this has a pronounced effect on the ultimate gaseous structure  \citep{backreaction}. We assumed a central stellar mass of 0.6 \msun, and a total disc mass of 0.3 \msun. The central star was represented by a sink particle  \citep{bate1995}, with accretion radius set to 1 au.

We set the initial inner and outer disc radii to 10 au and 300 au respectively. The surface density and sound speed profiles were set as $\Sigma_\mathrm{g}\propto R^{-1}$ and $c_\mathrm{s}\propto R^{-0.25}$ respectively. These properties are consistent with observed candidates for self-gravitating protoplanetary discs \citep{perezetal2016,dsharp,dsharpspirals}, and have been extensively used in previous modelling \citep{meruetal2017,tomidaetal2017,halletal2018,halletal2019}. We assumed a polytropic equation of state, and heating in the simulation is provided by shocks and pressure-volume ($P$ d$V$) work. The disc was initially set as stable to self-gravitating spirals, such that the Toomre parameter  \citep{toomre1964},$Q$ is 
\begin{equation}
    Q = \frac{c_\mathrm{s}\kappa}{\pi\mathrm{G}\Sigma} \gtrsim 2
\end{equation}
everywhere in the disc, where $\kappa$ is epicyclic frequncy, which for a disc in Keplerian rotation is simply $\Omega=\sqrt{ \frac{\mathrm{G} M }{R^3}}$, G is the gravitational constant and $\Sigma$ is surface density.  We implemented ``$\beta$" cooling  \citep{gammie2001}, a simple cooling prescription where the cooling timescale, $t_\mathrm{c}$, is a linear function of the dynamical timescale, such that $t_\mathrm{c} = \beta t_\mathrm{dyn}$. The dynamical timescale is simply the rotation period, $\frac{2\pi}{\Omega}$, and we set $\beta = 15$. We evolved the disc for several outer orbital periods. The spatial distribution of a dust grain of size $a_i$ and density $\rho_i$ is determined by its Stokes number  \citep{birnstieletal2010},
\begin{equation}
St = \frac{\pi a_i\rho_i}{2 \Sigma_\mathrm{g}}.
\end{equation}
The velocity of dust relative to the gas depends on $St$. For $St\ll1$, the dust is well-coupled and follows the gas drag. 
%RD: just to be accurate; thought they still responde to gas gravity
For $St \gg 1$, the dust is decoupled and does not respond to the gas. The maximum relative velocity occurs for $St=1$, resulting in particle trapping in the disc as density gradients peak inside spiral arms \citep{riceetal2004}. In our simulations, we see efficient trapping for particles $\gtrsim$mm. The model surface density is shown in Figure \ref{fig:splash}, along with two example grain sizes.\\~\\

\subsection{Thermal disc structure.}
\label{subsec:mcfost}

\begin{figure*}[t!]
  \includegraphics[width=\linewidth]{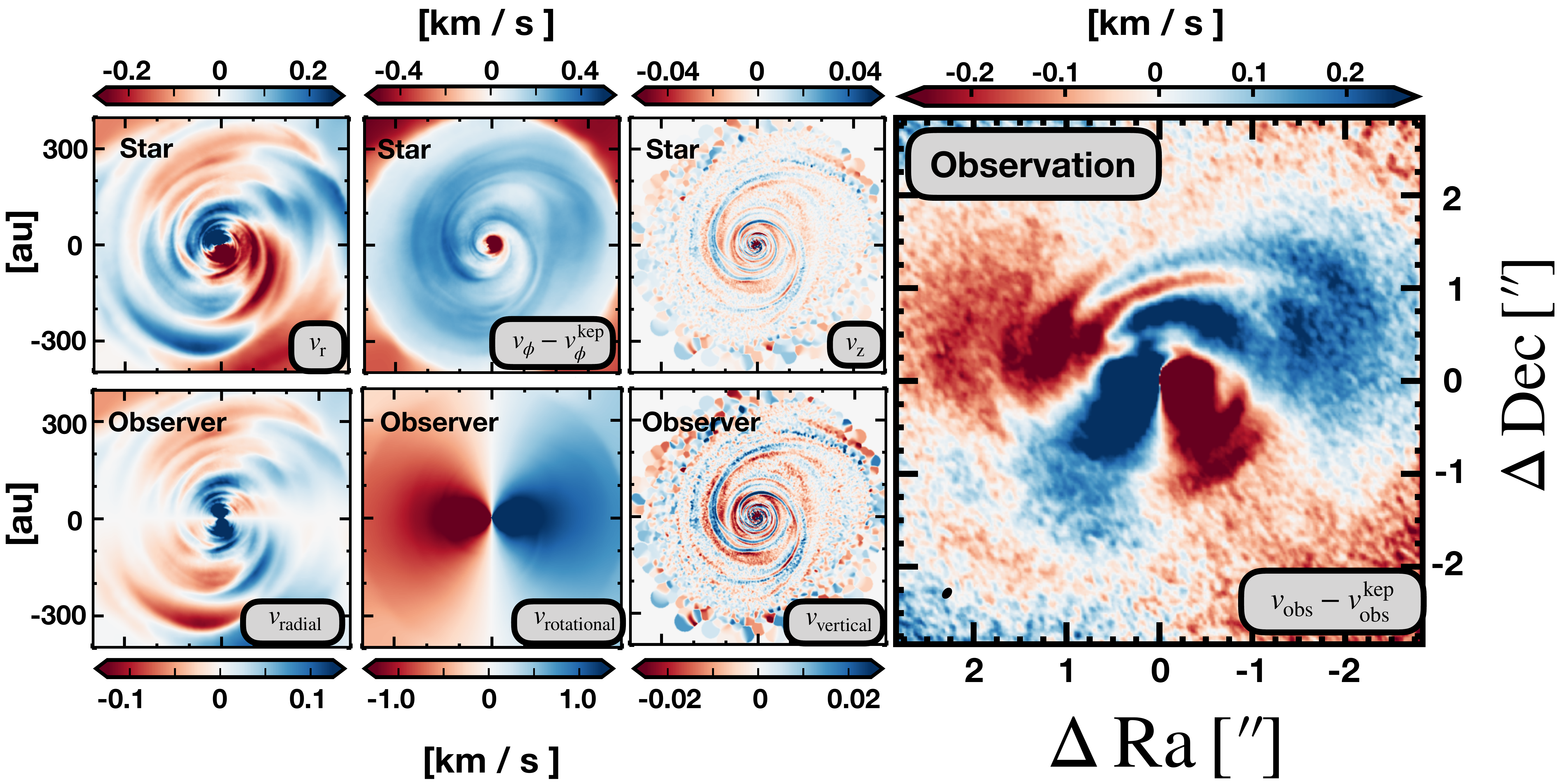}
\caption{\textbf{Actual and observed velocity fields.}Top 3 panels on the left show velocity elements in star reference frame. In Keplerian rotation $v_r=0$. Top center shows rotation is super-Keplerian throughout most of the disc. $v_\mathrm{z}$ is most positive in the density peak of spiral arm, and most negative in the inter-arm region. Bottom 3 panels show the velocity components in frame of observer at inclination of $30\degree$. End right panel is the total observed velocity field (also known as moment-1), with the projected Keplerian velocity field ($v_\phi= [\mathrm{G}M_*/r]^{\frac{1}{2}} ) $
subtracted. The velocity deviation, clearly observed as finger-like structure at $\sim$ [0\arcsec{},1\arcsec{}] pushes additional emission into a given velocity channel, responsible for the GI wiggle circled in Figure \ref{fig:channelmaps}.\label{fig:moment1}} 
\end{figure*}

We used the Monte Carlo radiative transfer MCFOST code \citep{mcfost1,mcfost2}  to compute the disc thermal structure and synthetic $^{13}$CO $J=3\rightarrow 2$ line maps. We assumed $T_\mathrm{gas} = T_\mathrm{dust}$, and used $10^8$ photon packets to calculate $T_\mathrm{dust}$. We also assumed that the $^{13}$CO molecule is in local thermodynamic equilibrium (LTE). It is reasonable to assume LTE for low-J lines since CO density is above the critical density for collisions to dominate over radiation. We set the $^{13}$CO abundance equal to $7\times 10^{-7}$ relative to the local H$_2$. The parameters for the central star were set to match those of a typical self-gravitating protostellar disc candidate, the Elias 2-27 system \citep{andrews2009}, with temperature $T=3850$ K, $M = 0.6$ \msun and $R_* = 2.3$ R$_\odot$. 

The SPH density structure underwent Voronoi tesselation such that each SPH particle corresponds to an MCFOST cell. The dust composition was assumed to be a mixture of silicate and amorphous carbon  \citep{drainelee1984} and optical properties were calculated using the Mie theory. We used a grain population with 100 logarithmic bins ranging in size from  0.03 $\mu$m to 4 mm. At each position in the model, the dust density of a grain size $a_i$ was obtained by interpolating from the SPH dust sizes. We assumed that grain sizes smaller than half the smallest SPH grain size (so 0.5 $\mu$m) are perfectly coupled to the gas distribution. We assume there are no grains larger than those present in the SPH simulation. The dust size distribution was normalised by integrating over all grain sizes, where a power-law relation between grain size $a$  and number density of dust grains $n(a)$ was assumed such that d$n(a)\propto a^{-3.5}$\,d$a$. \\~\\

%\subsection{$^{13}$CO channel map and moment-1 velocity field.}
\subsection{$^{13}$CO channel map and object velocity field} 
\label{subsec:obs}

The system was synthetically observed at a distance of 140 pc and inclination of 30 $\degree$. We assumed a turbulent velocity of 0.05 km s$^{-1}$. $^{13}$CO maps were generated at a Hanning-smoothed spectral resolution of 0.03 km s$^{-1}$, and then convolved with a beam of size 0.11\arcsec{} $\times$ 0.07\arcsec{}, with a position angle of $-38\degree$, matching recent ALMA observations that had the spectral and spatial resolution to kinematically detect a planet  \citep{pintenature}. Since we do not aim to perform a detailed fitting to any particular observation, we show our synthetic maps with fully sampled $uv$-plane. It has already been determined that this process does not affect observational results compared to more sophisticated analysis, and does indeed provide a good approximation for comparing models to data  \citep{pintenature}. The resulting channel map is shown in Figure \ref{fig:channelmaps}.

\begin{figure*}[t!]
  \includegraphics[width=\linewidth]{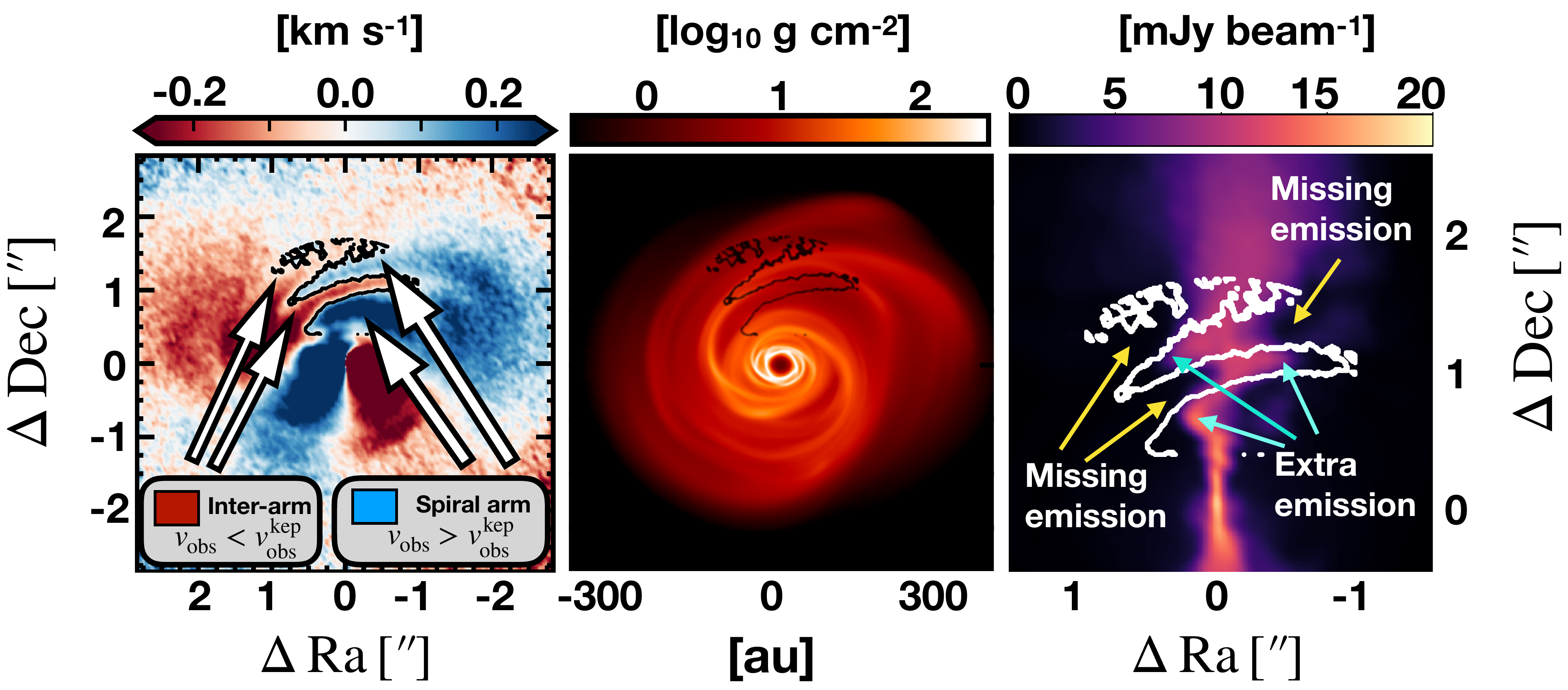}
\caption{\textbf{Relation between underlying structure and observed emission}. Left panel shows observed velocity relative to Keplerian rotation, with the line $v_\mathrm{obs} - v_\mathrm{obs}^{kep} = 0$ drawn in black. Center panel shows surface density calculated, calculated by integrating along the line of sight at an observed inclination angle of 30$\degree$. The line $v_\mathrm{obs} - v_\mathrm{obs}^{kep} = 0$ is shown in black. Faster emission relative to Keplerian comes from the density peak of spiral arm. Slower emission is from inter-arm region. Right panel is zoomed in velocity centroid (top center Figure \ref{fig:channelmaps}) showing how this affects observed emission.\label{fig:annotated}}
%RD: the right panel is the channel map middle panel in figure 2? Perhaps this could be made clear.
\end{figure*}

The velocity field of the simulation is shown in the leftmost panels of  Figure \ref{fig:moment1}. We determined the observed velocity field of this observation by calculating the intensity weighted average velocity of the emission line profile. This is also known as a ``moment-1'' map, and is obtained through
\begin{equation}
\label{eq:moment1}
    \langle v \rangle = \frac{\int^\infty_{-\infty} v I(v) \mathrm{d}v}{ \int^\infty_{-\infty} I (v) \mathrm{d} v},  
\end{equation}
where the denominator is simply the integrated line intensity. We then calculated equation \ref{eq:moment1} for the case where velocities were set to exactly Keplerian, and subtracted this result from the original velocities. The resulting observed velocity is shown in the rightmost panel of Figure \ref{fig:moment1}.

%and clearly shows long, finger like substructure in deviation from Keplerian rotation. These ``fingers'' essentially add and subtract  emission from each channel, causing the ``GI wiggle''. \\~\\

\section{Results}
Our results show that GI, unlike an embedded protoplanet, causes deviations from Keplerian rotation throughout the disc, resulting in velocity ``kinks'' across the entire radial and azimuthal extent of the disc. This is shown in the channel maps in Figure \ref{fig:channelmaps}. We call these kinks the ``GI wiggle''. 

%We measure the observed deviation from Keplerian rotation in the intesity-weighted average velocity (moment-1) map of the object, shown in the rightmost panel of Figure \ref{fig:moment1}. Long, interlocking, finger-like structure is present in the observed velocity. These features provide key evidence that a spiral is due to gravitational instability. 

We circle the GI wiggle at the systemic velocity, however it is clearly seen at all velocities in the disc. Unlike a planet-induced perturbation, which results in a localized kink, there are multiple kinks in the Keplerian cone. In the bottom row of Figure \ref{fig:channelmaps}, the velocity of the hydrodynamics simulation was set to equal exactly Keplerian, and synthetic line maps were generated in the exact same way as for the self gravitating case. There is no observed substructure in this case, which demonstrates that it is velocity perturbation, rather than perturbation of density structure, which is the cause of this GI wiggle. 

The velocity field of the self-gravitating disc is shown in Figure \ref{fig:moment1}. The top left panel shows the radial velocity in the reference frame of the star, top center panel shows the deviation from Keplerian rotation (where we define $v_\phi^\mathrm{kep}=({\mathrm{G}M_*/r})^{\frac{1}{2}}$ in  azimuthal velocity, and top right shows $z$-component of velocity (where positive $z$ is out of the page) all calculated directly from the hydrodynamics simulation. In Keplerian rotation, $v_{\mathrm{r}}=0$ and $v_\phi - v_\phi ^{\mathrm{kep}} = 0$. Bottom panels show each contribution to the observed velocity $v_\mathrm{obs}$, given by 

\begin{equation}
%\begin{aligned}
    v_\mathrm{obs} = \underbrace{v_{\phi} \sin(i) \cos(\phi)}_{\rm rotational} + 
          \underbrace{v_{r} \sin(i) \sin(\phi)}_{\rm radial} + 
          \underbrace{v_{z} \cos(i)}_{\rm vertical} + v_{\rm systemic}.
    \label{eq:observed_vel}
%\end{aligned}
\end{equation}

The rightmost panel is the observed velocity field (moment-1), with observed Keplerian rotation subtracted. It clearly shows interlocking, finger-like structure between [0\arcsec{}, 0\arcsec{}] and [0\arcsec{},1.5\arcsec{}]. This deviation essentially pushes extra emission into adjacent channels at a given velocity, causing the GI wiggle seen in Figure \ref{fig:channelmaps}. We emphasise that perturbations do not create extra emission, but simply relocate emission in position-position-velocity space. 

Figure \ref{fig:moment1} shows that GI spirals have strong velocity perturbations across their azimuthal extent. This is why the GI wiggle is seen in multiple channels, as shown in Figure \ref{fig:channelmaps}. Planetary companions, on the other hand, create velocity deviations that are only strong enough to be detected in line emission close to the planet. Figure \ref{fig:moment1} also shows that GI spirals have multiple perturbations across radial extent of the disc, which is why the GI wiggle has multiple inflection points. 

It is difficult to directly relate the velocity structure in left and center panels of Figure \ref{fig:moment1} to the observed velocity, since the observed velocity, $v_\mathrm{obs}$, is the superposition of the projection of rotational, radial, and vertical components, as demonstrated in equation \ref{eq:observed_vel}.

Under the assumption of an azimuthally symmetric velocity distribution, these three velocity components are readily disentangled  \citep{teagueetal2019}, since they each have differing dependence on azimuthal angle $\phi$. However, as demonstrated by Figure \ref{fig:moment1}, the velocity distribution in a self-gravitating disc deviates strongly from azimuthal symmetry. 

%\ch{ARM PUSHES EMISSION IN FROM FASTER CHANNELS, INTER ARM PUSHES EMISSION IN FROM SLOWER CHANNELS}

We attribute these features to the underlying disc structure in Figure \ref{fig:annotated}. Left panel shows $v_\mathrm{obs} - v_\mathrm{obs}^{\mathrm{kep}}$ in the moment-1 map, where we have traced the line of $v_\mathrm{obs} - v_\mathrm{obs}^{\mathrm{kep}} = 0$ in the interlocking fingers. Center panel shows the projected surface density of the disc calculated at the observed inclination angle (30$\degree$), integrated along this line of sight. The location of $v_\mathrm{obs} - v_\mathrm{obs}^{kep} = 0$ is plotted in black. The strong perturbations in radial velocity, shown in Figure \ref{fig:moment1}, cause a radius of faster rotating material to have a higher velocity in the red-shifted side of the disc, but a lower velocity in the blue-shifted side.

The right panel of Figure \ref{fig:annotated} is a zoomed in version of the velocity centroid ($\Delta v = 0.0$ km s$^{-1}$) in Figure \ref{fig:channelmaps}. Over-plotted in white is the line $v_\mathrm{obs} - v_\mathrm{obs}^{kep} = 0$. At the base of each ``finger'' there is missing emission, which has been ``stolen'' by the adjacent velocity channel. For the $v_{\mathrm{obs}}< v_\mathrm{obs}^{\mathrm{kep}}$ fingers, it has been ``stolen'' by the slower velocity channel adjacent to the centroid. For the $v_{\mathrm{obs}}> v_\mathrm{obs}^{\mathrm{kep}}$ fingers, it has been stolen by the faster velocity channel adjacent to the centroid. At the tip of each finger, extra emission is present by the same mechanism. 

\begin{figure*}[t!]
  \includegraphics[width=1.0\linewidth]{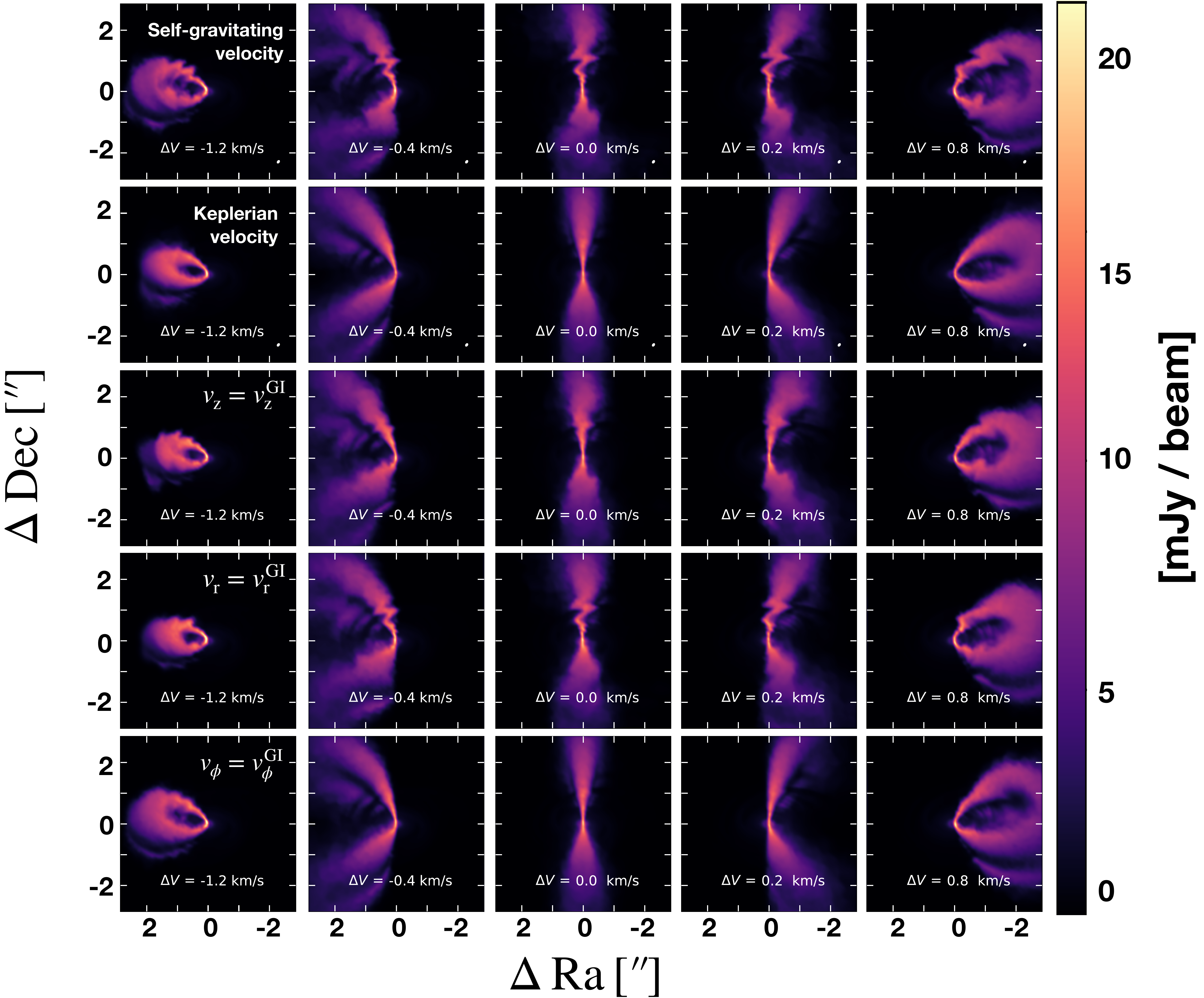}
\caption{ \textbf{Emission for GI and Keplerian velocity components.} Top two rows show predicted emission for GI and for exactly Keplerian rotation. 
Center row shows GI $z$-velocity component, with $v_\mathrm{r}$ and $v_\phi$ set to Keplerian, and bottom two rows show $v_\mathrm{r}$ and $v_\phi$ GI velocity with all other velocity components set to Keplerian.
Perturbations in $v_\mathrm{r}$ are strongest and are seen throughout disc azimuth and radius. Perturbations in $v_\mathrm{z}$ are weaker but visible throughout azimuth and radius. Perturbations in $v_\phi$ only seen at specific azimuths.  \label{fig:channelmaps_vGI}}
\end{figure*}

%\captionsetup[figure]{labelfont={bf},labelformat={default},labelsep=colon,name={Supplementary Figure}}

\begin{figure*}[t!]
\centering
  \includegraphics[width=0.9\linewidth]{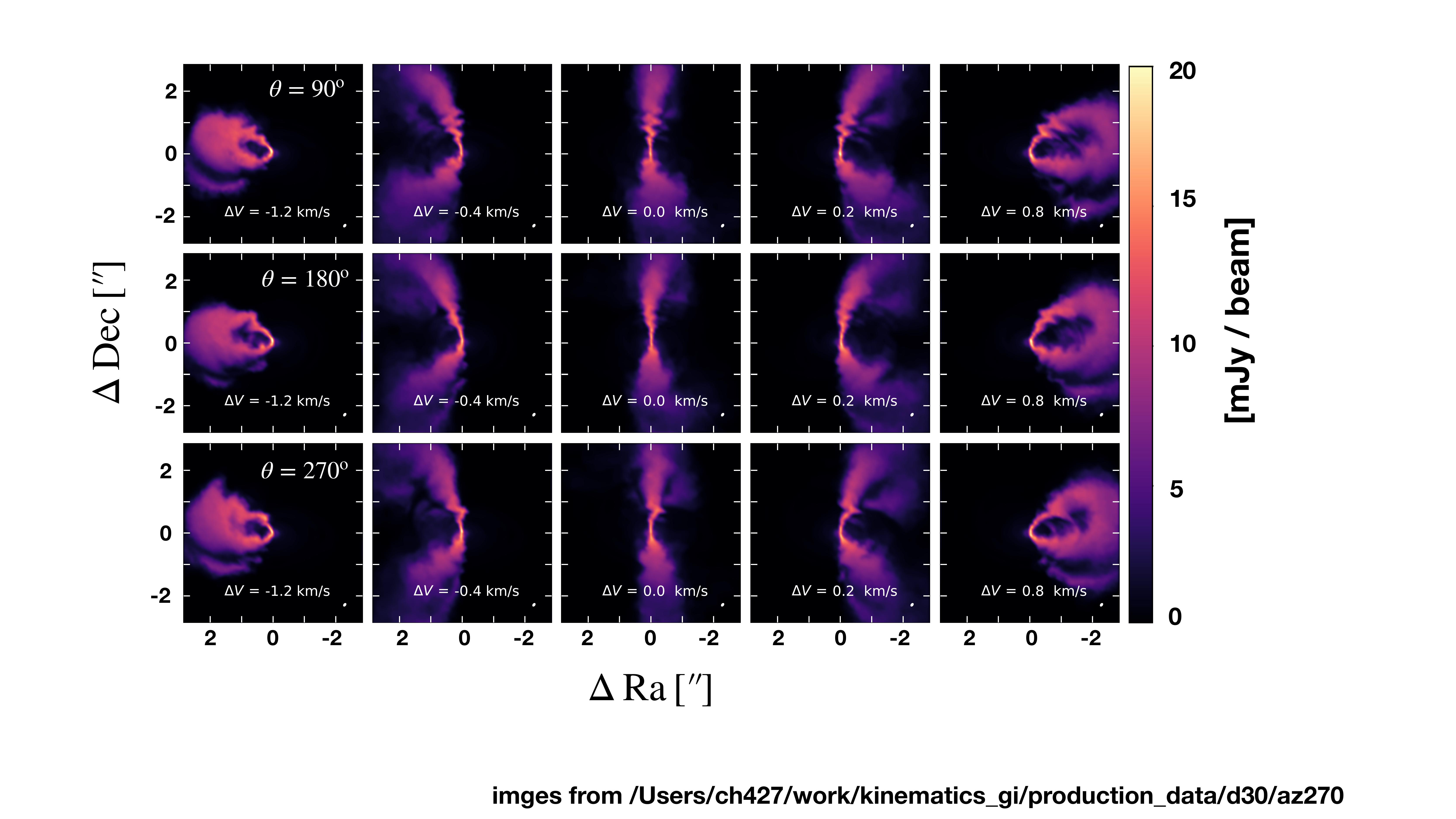}
\caption{\textbf{Emission for different viewing angles.} All discs are observed at an inclination of 30$\degree$, with azimuthal viewing angle varying as $\theta = 90\degree, 180\degree $ and $270\degree$. The GI wiggle is visible at all viewing angles, but does vary in number of observed inflection points and amplitude of wiggle.}
\label{fig:channelmaps_azimuths}
\end{figure*}

\begin{figure*}[t!]
\centering
  \includegraphics[width=0.9\linewidth]{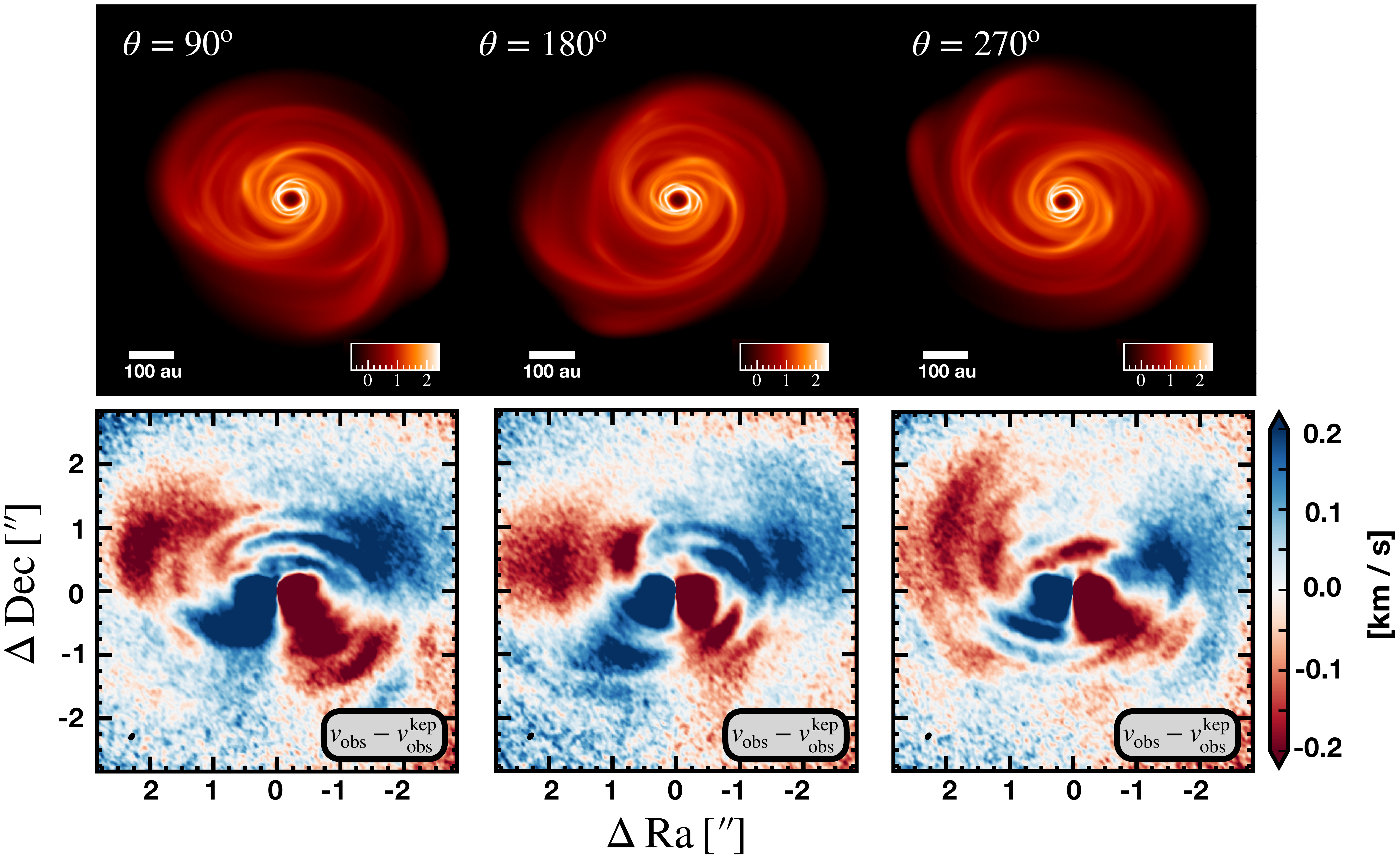}
\caption{ \textbf{Column density and velocity field for different viewing angles.} Top panel shows integrated column density in units of log$_{10}$ g cm$^{-2}$, at an inclination of 30$\degree$ and varying viewing angle. Bottom panels show the observed velocity field with Keplerian rotation profile subtracted. Strong interlocking fingers are only seen in $\theta=90\degree$, but non-axisymmetry is present at all viewing angles.
\label{fig:moment1_azimuths}}
\end{figure*}

\subsection{Contribution of velocity components}
The six leftmost panels in Figure \ref{fig:moment1} show that GI discs differ significantly from Keplerian rotation in $v_{\mathrm{r}}$, $v_{\mathrm{\phi}}$ and $v_\mathrm{z}$. To determine which velocity component deviations contribute most strongly to the GI wiggle, we repeat the procedure outlined above for three additional cases:
\begin{enumerate}
    \item $v_\mathrm{z}$ as the perturbed GI velocity, $v_\mathrm{r} = 0$ and $v_\phi = v_\mathrm{Keplerian}$. 
    \item $v_\mathrm{r}$ as the perturbed GI velocity, $v_\mathrm{\phi} = v_\mathrm{Keplerian}$ and $v_\mathrm{z} = 0$. 
    \item $v_\mathrm{phi}$ as the perturbed GI velocity, $v_\mathrm{r} = 0$ and $v_\mathrm{z} = 0$ set to Keplerian rotation, 
    
\end{enumerate}
where $v_\mathrm{r}=v_\mathrm{z}=0$ in Keplerian rotation. This is shown in Figure \ref{fig:channelmaps_vGI}.  We determine that the perturbations in $v_\mathrm{r}$ are the strongest contributors to the GI wiggle, and are visible throughout azimuthal and radial disc extent. The $v_\mathrm{z}$ perturbations are also visible throughout azimuthal extent, but are of smaller observed amplitude. The perturbations in $v_\phi$ are not seen throughout azimuthal extent. \\~\\

\subsection{Robustness to viewing angle}
We ensure that the detection of the GI wiggle is robust to the geometry of the observation by generating synthetic observations at azimuthal viewing angle $\theta = 90\degree, 180\degree$ and $270\degree$. We show the resulting channel maps in Figure \ref{fig:channelmaps_azimuths}. The GI wiggle is observed at all radii and azimuths in the disc for all viewing positions in the channel maps. It does, however, have some variation both in amplitude and number of inflection points.

Self-gravitating discs, by nature, are not axisymmetric. Therefore, some variation with viewing angle is expected. We plot integrated column density at an inclination of 30$\degree$ in the top panels of Figure \ref{fig:moment1_azimuths}, for the viewing angles $\theta$ of 90, 180 and 270 degrees. The corresponding moment-1 maps, with Keplerian background subtracted, are shown below. For $\theta=90\degree$, interlocking finger-like structure is observed, while this is not present for $\theta = 180\degree, 270\degree$. However, in all cases, the velocity field shows clear deviation from axisymmetry and from Keplerianity. We conclude that the GI wiggle in the channel maps is the strongest kinematic signature for GI, and most likely to be detected, and interlocking structure in the moment-1 deviation from Keplerian is a strong secondary signal that may depend on viewing angle.\\~\\

\section{Conclusion}
We have demonstrated that velocity perturbations due to gravitational instability, in a disc imaged at 140 pc, have a clear kinematic signature that is detectable with current ALMA capabilities: a spatial resolution of $\sim 0.1$ \arcsec{} and a spectral resolution of $0.03$ km s$^{-1}$. Although planetary in origin, analysis of archival ALMA data by \citet{pinteetal2018} recovered velocity perturbations of similar amplitudes in the protoplanetary disc HD 163296, with a total observing time of 4.7 hours, and 2.5 hours on the science target \citep{isellaetal2016}. 

%ALMA

%spatial and spectral resolution of current observations  \citep{pintenature}. 

Unlike spirals caused by embedded planets, GI spirals do not cause a localised velocity deviation. They perturb the velocity throughout the disc, resulting in sustained ``GI wiggles'' that are visible at all disc radii and all azimuthal angles. Furthermore, they may leave clear, finger-like signatures in the observed velocity field of the system, particularly pronounced when a Keplerian rotation profile is subtracted.  The detection of the GI wiggle would provide strong evidence for the existence of gravitational instability in protoplanetary discs.

%  \subsection{Data Availability} The datasets generated and analysed during the current study are available from the corresponding author upon reasonable request.
%  \subsection{Code Availability} The hydrodynamics SPH code PHANTOM is publicly available from \url{https://bitbucket.org/danielprice/phantom}. MCFOST is currently available under request from Christophe Pinte (christophe.pinte@monash.edu)and will be made open-source soon. Some figures were generated with SPLASH  \citep{splash} (\url{http://users.monash.edu.au/~dprice/splash/}), which is open source.\\~\\

\section{Acknowledgements}
CH is a Winton Fellow and this research has been supported by Winton Philanthropies / The David and Claudia Harding Foundation. CH, BV and TP have received funding from the European Union’s Horizon 2020 research and innovation programme under the Marie Sklodowska-Curie grant agreement No 823823 (RISE DUSTBUSTERS project). C.P. acknowledges funding from the Australian Research Council via FT170100040 and DP180104235. RT acknowledges support from the Smithsonian Institution as a Submillimeter Array (SMA) Fellow. R.D. acknowledges support from the Natural Sciences and Engineering Research Council of Canada. RDA acknowledges support from the European Research Council (ERC) under the European Union’s Horizon 2020 research and innovation programme (grant agreement No 681601). This research used the ALICE2 High Performance Computing Facility at the University of Leicester. This research also used the DiRAC Data Intensive service at Leicester, operated by the University of Leicester IT Services, which forms part of the STFC DiRAC HPC Facility (\url{www.dirac.ac.uk}). The equipment was funded by BEIS capital funding via STFC capital grants ST/K000373/1 and ST/R002363/1 and STFC DiRAC Operations grant ST/R001014/1. DiRAC is part of the National e-Infrastructure.  We would like to thank Daniel Price for his publicly available SPH plotting code SPLASH  \citep{splash}, which we have made use of in this paper.

 %\item[Author Contributions] CH performed the hydrodynamics simulation, radiative transfer calculation, synthetic observation and data analysis. CH and RD wrote the manuscript. RD and RT made suggestions for data analysis. JT assisted with a portion of the data analysis. CP wrote the radiative transfer code MCFOST used to calculate molecular line maps. TP and BV provided valuable discussion at the beginning of the project. RT, RD, RDA, CP and GL provided comments on the data and the manuscript. 
 %\item[Competing Interests] The authors declare that they have no
%competing financial interests.
% \item[Correspondence] Correspondence and requests for materials
%should be addressed to Cassandra Hall (email: cassandra.hall@le.ac.uk).

\bibliographystyle{aasjournal}
\bibliography{bib}{}

%Set these things if want to count differently the supplementary figures
%\setcounter{figure}{0}

%\captionsetup[figure]{labelfont={bf},labelformat={default},labelsep=colon,name={Supplementary Figure}}

\end{document}